\documentclass[fleqn,12pt,twoside]{article}
\usepackage{espcrc1}

\usepackage{graphicx}
\usepackage[figuresright]{rotating}

\renewcommand{\vec}[1]{\mbox{\boldmath $#1$}}

\newcommand{\AmS}{{\protect\the\textfont2
  A\kern-.1667em\lower.5ex\hbox{M}\kern-.125emS}}

\title{Continuum QRPA response for deformed neutron-rich nuclei}

\author{K. Hagino\address{Yukawa Institute for Theoretical Physics, 
Kyoto University, Kyoto 606-8502, Japan}
\address{Institut de Physique Nucl\'eaire, IN2P3-CNRS, 
Universit\'e Paris-Sud, F-91406 Orsay Cedex, France},
Nguyen Van Giai\addressmark, 
H. Sagawa\address{Center for Mathematical Sciences, University of Aizu,
Aizu-Wakamatsu, 
Fukushima 965-8560, Japan}}
       
\begin{document}

\maketitle

\begin{abstract}

We discuss properties of the quadrupole collective excitation of 
the deformed neutron-rich nucleus $^{38}$Mg within the framework of 
quasi-particle random phase approximation (QRPA). 
We first solve the coupled-channels equations to obtain 
the single-particle levels, and construct the ground 
state by treating the 
pairing correlations in the BCS approximation. 
We then solve the QRPA equation using the response function formalism, by 
including the continuum spectra with the box dicscretization method. 
We show that the collectivity of the gamma vibration (the lowest 
$K^\pi=2^+$ mode) 
is significantly enhanced if protons and neutrons have different 
deformations. 
We also discuss an attempt towards full continuum QRPA calculations for 
deformed nuclei. 

\end{abstract}

\section{Introduction}

Without an exception, 
stable nuclei with a neutron number $N$=20 or 28, as well as their 
neigbouring nuclei, have a spherical intrinsic shape. 
Heavy neutron-rich nuclei, however, often exhibit deformed characters 
even around the shell closure. Typical examples include 
$^{32}$Mg \cite{M02}, $^{34}$Mg \cite{Y01,I01}, and $^{44}$S \cite{S02}. 
This phenomenon is often interpreted as a disappearance of spherical 
magic shells \cite{WBB90,O01}. 

In this contribution, we investigate the excitation structures of such 
{\it deformed} neutron-rich nuclei within the framework of 
quasi-particle random phase approximation (QRPA). 
For nuclei far from 
the $\beta$ stability line, it has been well recognised that both the 
pairing interaction and the couplings to the continuum spectra play an 
important role. The most robust and consistent framework to 
deal with these effects on the ground state properties is the 
Hartree-Fock-Bogoliubov (HFB) method \cite{D96,G01}. 
The continuum QRPA framework based on the HFB approximation has also 
been developed recently \cite{M01,K02}. 
A problem with the HFB method, however, is that it can be computationally 
very demanding. For this reason, the application of the continuum QRPA 
on top of the HFB ground state 
has been limited only to spherical systems so 
far \cite{M01,K02}. 
In this paper, we avoid this difficulty by using the BCS approximation. 
This is an approximation to the HFB method, where one takes only 
time-reversed pairs in the Bogoliubov transformation. 
With this approximation, the particle-hole and the particle-particle 
channels are decoupled completely (for a density-independent 
pairing force at least) 
both in the ground state and in the QRPA calculations. 
Notice that 
the BCS method has been shown a reasonable approximation to the HFB, 
within the same treatment for the continuum states, 
even for nuclei close to the neutron drip line \cite{G01,SGL00}. 

Another difficulty related to a deformed QRPA calculation is that the 
configuration space can be quite large. In the past, QRPA calculations 
for a deformed nucleus have been done either with a simple separable 
interaction \cite{SM84} or with a drastic truncation of the 
configuration space \cite{ZSP78}. In this paper, we present a 
formalism which uses the response function in the coordinate 
space  \cite{HS01}, 
which is most suitable to a zero-range interaction, eliminating 
these limitations. 
Together with the BCS approximation for the pairing interaction, 
we will demonstrate below that a deformed QRPA calculation now becomes 
feasible within a reasonable computation time. 

\section{Formalism}

\subsection{Coupled-channels approach to single-particle spectra}

Let us consider a single-particle motion in a deformed mean-field 
potential, 
\begin{equation}
\hat{V}(\vec{r})=V_{\rm cent}(\vec{r})+
\nabla(V_{\rm ls}(\vec{r}))
\cdot(-i\nabla\times\vec{\sigma}), 
\label{pot}
\end{equation}
with
\begin{eqnarray}
V_{\rm cent}(\vec{r})&=&V_0(r)-R\,\beta\,\frac{dV_0(r)}{dr}
\,Y_{20}(\hat{\vec{r}}) + 
\left(V_C(r) + \frac{3(Z-1)}{5}\frac{R^2}{r^3}\beta\,Y_{20}(\hat{\vec{r}})\right)
(1-\tau_z), \label{pot0}\\
V_{\rm ls}(\vec{r})&=&V_{\rm so}(r)-R_{\rm so}\,\beta\,
\frac{dV_{\rm so}(r)}{dr}\,Y_{20}(\hat{\vec{r}}). 
\label{potls}
\end{eqnarray}
Here, $\beta$ is the deformation parameter. For simplicity, 
we take into account only the linear order of $\beta$ in the potential. 
We have also assumed an axial symmetry, where both the parity $\pi$ and 
the spin projection $K$ onto the $z$ axis 
are conserved. 

In order to obtain the single-particle levels and 
corresponding wave functions, we solve 
the coupled-channels equations for both negative and positive 
energy states \cite{HG03}. 
This method has an advantage that the correct scattering boundary conditions  
can be easily implemented for the positive energy solutions, 
that is important in a full continuum 
calculation for neutron-rich nuclei. 
In this paper, for the sake of simplicity,  
we instead discretize the 
positive energy spectra within the same framework 
by putting the nucleus in a large 
spherical box, as was recently demonstrated in Ref. \cite{HG03}, and leave 
the full continuum treatment as a future work (see sect. 4). 

The coupled-channels method has sometimes a difficulty in finding all 
solutions in a deformed potential, especially when there are two states 
close to each other in energy. One can easily miss a few 
eigenvalues in a search routine. We have overcome this difficulty by 
introducing the generalized wave function node for a multi-channel 
system, which was originally proposed by Johnson \cite{J78}. 
We have checked the efficiency of this method by applying it to a spherical 
system. We have also computed the Nilsson diagram of $^{238}$U and have 
confirmed that all the levels obtained were smoothly 
connected to the spherical levels. Evidently, the coupled-channels method 
with the generalized wave function node efficiently works even for 
heavy nuclei, where there are lots of intruder levels. 

Using the single-particle levels obtained in this way, we then solve the 
BCS gap equation. 
As we emphasized in the introduction, we use the BCS approximation rather 
than the HFB approximation. The applicability of the BCS approximation 
for neutron-rich nuclei has been well tested in Ref. \cite{G01,SGL00}. 
In this paper, we particularly use the density-dependent zero-range 
force for the pairing interaction. We will specify the parameters in the 
next section. 

\subsection{Response function formalism for deformed QRPA} 

The elements of the QRPA matrices have been given in the textbook of 
Rowe \cite{R68}: 
\begin{eqnarray}
A_{\alpha\beta,\gamma\delta}&=&
(E_\alpha+E_\beta)\delta_{\alpha\beta,\gamma\delta} 
+\langle \alpha\bar{\delta}|v_{ph}|\bar{\beta}\gamma\rangle\, 
(u_\alpha v_\beta + v_\alpha u_\beta)(u_\gamma v_\delta + v_\gamma u_\delta)
\nonumber \\
&&+
\langle \alpha\beta|v_{pp}|\gamma\delta\rangle\,
(u_\alpha u_\beta u_\gamma u_\delta + v_\alpha v_\beta v_\gamma v_\delta), 
\label{qrpa-a}
\\
B_{\alpha\beta,\gamma\delta}&=&
\langle \alpha\gamma|v_{ph}|\bar{\beta}\bar{\delta}\rangle\, 
(u_\alpha v_\beta + v_\alpha u_\beta)(u_\gamma v_\delta + v_\gamma u_\delta) 
\nonumber \\
&&-\langle \alpha\beta|v_{pp}|\bar{\gamma}\bar{\delta}\rangle\,
(u_\alpha u_\beta v_\gamma v_\delta + v_\alpha v_\beta u_\gamma u_\delta),
\label{qrpa-b}
\end{eqnarray}
where $E_\alpha$ is the quasi-particle energy and $\bar{\alpha}$ is 
the time reversed state of $\alpha$. 
In writing down these equations, we have used the identity, 
$\langle \alpha\bar{\delta}|v|\bar{\beta}\gamma\rangle = 
-\langle \alpha\bar{\gamma}|v|\bar{\beta}\delta\rangle$, which 
follows from the property of the time-reversal operator. 
Notice that the particle-particle matrix elements vanish in 
the BCS approximation unless $\beta=\bar{\alpha}$ and 
$\delta=\bar{\gamma}$ in eqs. (\ref{qrpa-a}) and (\ref{qrpa-b}). 
On the other hand, the particle-hole contribution is finite only if 
$\beta\neq\bar{\alpha}$ and 
$\delta\neq\bar{\gamma}$ due to the BCS phase convention, 
$v_{\bar{\alpha}}=-v_\alpha$ and $u_{\bar{\alpha}}=u_\alpha$. Therefore, 
the particle-particle and the particle-hole channels are decoupled 
completely in the BCS approximation, not only in the ground state but also in 
the QRPA calculations. 

Retaining only the particle-hole channel, one can write down the QRPA 
equations in the following form: 
\begin{eqnarray}
0&=&(E_{\alpha}+E_{\beta}-\omega)X_{\alpha\beta}
+\int\,d\vec{r}d\vec{r}'\,D_{\alpha\bar{\beta}}(\vec{r})\,
v_{ph}(\vec{r},\vec{r}')\,T(\vec{r}'), \\
0&=&(E_{\alpha}+E_{\beta}+\omega)Y_{\alpha\beta}
+\int\,d\vec{r}d\vec{r}'\,D^*_{\alpha\bar{\beta}}(\vec{r})\,
v_{ph}(\vec{r},\vec{r}')\,T(\vec{r}').
\end{eqnarray}
Here, we have explicitly 
expressed the matrix elements of the particle-hole interaction 
in the integral form. 
Using the single-particle wave function 
$\phi_\alpha(\vec{r})$, 
$D_{\alpha\bar{\beta}}(\vec{r})$ and 
$T(\vec{r})$ are defined by  
\begin{eqnarray}
D_{\alpha\bar{\beta}}(\vec{r})&=&
(u_\alpha v_\beta + v_\alpha u_\beta)
\phi^*_\alpha(\vec{r})\phi_{\bar{\beta}}(\vec{r}),\\
T(\vec{r})&=&\sum_{\gamma\delta}
(D^*_{\gamma\bar{\delta}}(\vec{r})X_{\gamma\delta}
+D_{\gamma\bar{\delta}}(\vec{r})Y_{\gamma\delta}),
\end{eqnarray}
respectively. From these equations, one obtains the generalised (Q)RPA 
dispersion relation \cite{GSV98}, 
\begin{equation}
0=
\int\,d\vec{r}'\,\left[\delta(\vec{r}-\vec{r}')
-\int\,d\vec{r}''\,\Pi_0(\vec{r},\vec{r}'')\,
v_{ph}(\vec{r}'',\vec{r}')\right]\,T(\vec{r}')
\equiv 
\int\,d\vec{r}'\,W(\vec{r},\vec{r}')\,T(\vec{r}'),
\end{equation}
where the unperturbed response function is given by 
\begin{equation}
\Pi_0(\vec{r},\vec{r}'')
=-\sum_{\alpha\beta}\left(
\frac{D^*_{\alpha\bar{\beta}}(\vec{r})D_{\alpha\bar{\beta}}(\vec{r}')}
{E_{\alpha}+E_{\beta}-\omega-i\eta}
+\frac{D_{\alpha\bar{\beta}}(\vec{r})D^*_{\alpha\bar{\beta}}(\vec{r}')}
{E_{\alpha}+E_{\beta}+\omega+i\eta}\right).
\label{Pi0}
\end{equation}

The response of the system to an external field can be calculated using 
the QRPA response function defined by \cite{HS01}
\begin{equation}
\Pi(\vec{r},\vec{r}')=\int\,d\vec{r}''\,W^{-1}(\vec{r},\vec{r}'')\,
\Pi_0(\vec{r}'',\vec{r}').
\label{QRPA}
\end{equation}
For axially symmetric wave functions $\phi_\alpha$, the response 
function can be decomposed into multipoles as
\begin{equation}
\Pi(\vec{r},\vec{r}')=\sum_{L,L'}\sum_K 
\frac{1}{rr'}\,\Pi_K(rL,r'L')\,
Y_{LK}(\hat{\vec{r}})Y^*_{L'K}(\hat{\vec{r}}').
\end{equation}
We assume a zero range particle-hole interaction of the form 
\begin{equation}
v_{ph}(\vec{r},\vec{r}')
=\left[\sum_{\lambda: even} v_\lambda(r)Y_{\lambda 0}(\hat{\vec{r}})
\right]\,\delta(\vec{r}-\vec{r}').
\label{interaction}
\end{equation}
Then, the QRPA response function (\ref{QRPA}) is obtained by 
solving the Bethe-Salpeter equation for each $K$, 
\begin{equation}
\Pi_K(rL,r'L')=
\Pi^{(0)}_K(rL,r'L')
+\sum_{L''}\int\,dr''\,\left[\Pi^{(0)}v\right]_K(rL,r''L'')\,
\Pi_K(r''L'',r'L'),
\end{equation}
where
\begin{equation}
\left[\Pi^{(0)}v\right]_K(rL,r'L')
\equiv \sum_\lambda\sum_{L''}
\Pi^{(0)}_K(rL,r'L'')v_\lambda(r')\langle Y_{L''K}|Y_{\lambda 0}|
Y_{L'K}\rangle. 
\end{equation}
The strength function $S$ for an external field 
$V_{\rm ext}(\vec{r})=f(r)Y_{LK}(\hat{\vec{r}})$ 
is then obtained in the usual way as,
\begin{eqnarray}
S(\omega)&=&
-\frac{1}{\pi}\,Im\int\,d\vec{r}d\vec{r}'\,
V^*_{\rm ext}(\vec{r})\,\Pi(\vec{r},\vec{r}')\,
V_{\rm ext}(\vec{r}')   \\
&=&
-\frac{1}{\pi}\,Im\int\,r^2dr\,r'^2dr'\,
\frac{1}{rr'}\Pi_K(rL,r'L)f(r)f(r'),
\end{eqnarray}
where $Im$ denotes the imaginary part. 

A typical dimensionality of the QRPA response function 
$\Pi_K$ goes as follows. 
If one includes the radial mesh up to $r$=15 fm with 0.5 fm step, there 
are 30 points for the $r$ integrals. For the interaction given by 
eq. (\ref{interaction}), only those $L$ with the same parity are coupled 
to each other, and generally the convergence is rather quick. Therefore, for 
the quadrupole response for example, one would include only $L=0,2$ and 4. 
By treating proton and neutron separately, the total dimension 
of the response function is 
30 (for $r$) $\times$3 (for $L$) $\times$2 (for $\tau_z$) = 180.  
This is much smaller than the number of 2 quasi-particle 
configurations, which can be more than several thousands for a 
deformed system.  

\section{Quadrupole collectivity of $^{38}$Mg}

We now apply the formalism to the quadrupole response of the 
$^{38}$Mg nucleus. 
We assume a Woods-Saxon form for the mean-field potential $V_0(r)$ and 
$V_{\rm so}(r)$ in eqs. (\ref{pot0}) and (\ref{potls}) with 
parameters given in Ref. \cite{BM69}. 
We take the deformation parameters $\beta$ from the results of 
HFB calculations with a Skyrme interaction by Terasaki {\it et al.} 
\cite{TFHB97}. For $^{38}$Mg, the deformation parameters are 
0.33 and 0.28 for protons and neutrons, respectively. 
Notice that the deformation 
parameters are different for protons and neutrons by an amount of 0.05. 
We use the same pairing interaction as in Ref. \cite{TFHB97} for our BCS 
calculations. This is a density-dependent zero-range interaction with 
an energy cut-off at 5 MeV above the Fermi energy. Although a smooth energy 
cut-off was employed in Ref. \cite{TFHB97}, we use a sharp energy cut-off 
for simplicity. 
With these parameters, we obtain $\lambda_p = -24.4$ MeV and 
$\lambda_n = -0.56$ MeV for the proton and the neutron Fermi energies, 
$\bar{\Delta}_p=0.0$ MeV and $\bar{\Delta}_n=1.50$ MeV for the proton and 
the neutron average pairing gaps. 
For the residual interaction $v_{ph}$, we use the 
SIII parameter set \cite{S3} of the Skyrme interaction 
with the Landau-Migdal 
approximation for the momentum dependent terms \cite{GSV98}. 
We include the single-particle levels up to 50 MeV in the calculations. 
The continuum states are discretized with the box discretization method 
with the box size of 15 fm. We have checked that the results are not altered 
by more than 10\% even if we use a larger box size. For single-particle states 
above the pairing cut-off energy, we simply take $v^2=0$ for the BCS 
occupation probabilities. 

Since our calculations are not self-consistent in the sense that the 
residual interaction is not related to the mean field potential, 
we renormalize the residual interaction as $f\cdot v_{ph}$ 
so that the Goldstone modes appear at zero excitation energy. 
For a spherical system, there is only one Goldstone mode for 
the center-of-mass motion, and one can use a single value for the 
renormalization factor in order to 
define the effective residual interaction and study 
several multipole responses. In contrast, for a deformed 
system, the center-of-mass motion splits into two excitation modes 
according to the $K$ quantum number ($K^\pi = 0^-$ and 1$^-$). Moreover, 
there is another Goldstone mode associated with the rotation motion 
of the whole 
nucleus. This motion carries $K^\pi = 1^+$. 
We find that the renormalization factors required to put these modes at 
zero energy are significantly different from each other for the 
configuration space mentioned above. The renormalization factors are 
found to be 1.00075 for $K^\pi=0^-$, 0.95585 for $K^\pi=1^-$, 
and 1.0300 for $K^\pi=1^+$. Since we are interested in only 
the quadrupole motion 
in this paper, our prescription is to use $f_{1^+}$ for the other 
$K$ quantum numbers, i.e., $K^\pi = 0^+$ and $2^+$. 
Note that the Goldstone mode for the number conservation does not appear 
in our calculation, 
since we include only the particle-hole channel in QRPA. 

Figure 1 shows the isoscaler (IS) quadrupole strength distribution 
of the $^{38}$Mg nucleus 
(the lower panel) in comparison with that of the 
$^{24}$Mg nucleus (the upper panel). We set $\eta=0.5$ MeV in eq. (\ref{Pi0}) 
in order to smear out the results for the presentation purpose. 
The dotted, dashed, 
and solid lines are for $K^\pi$=0$^+$, 1$^+$, and 2$^+$ modes, 
respectively. Since we use a density-dependent pairing interaction, there 
is a small coupling between the particle-hole and the particle-particle 
channels for the $K^\pi=0^+$ mode (notice that a time-reversed pair can 
form only a $K^\pi=0^+$ state) even in the BCS approximation, when 
the residual interaction is evaluated as the second derivative of the energy 
functional \cite{M01,K02}. We simply neglect this effect in this work. 
The $K^\pi$=1$^+$ and 2$^+$ modes are free from this approximation. 

In the figure, one observes the splitting of the giant quadrupole 
resonances (GQR) \cite{YLC99} in the energy region 
10 -- 20 MeV both in $^{24}$Mg and in $^{38}$Mg. 
In contrast to the stable nucleus 
$^{24}$Mg, we notice that the IS GQR is more fragmented in the 
neutron-rich nucleus $^{38}$Mg due to a large neutron contribution. 
We have also computed the neutron response, and found that the 
$K^\pi=2^+$ mode in the energy region 3 -- 10 MeV is almost a pure 
neutron mode. Another interesting result is that the collectivity 
of the lowest $K^\pi=2^+$ mode is increased considerably in 
$^{38}$Mg. This is due both to the pairing effect \cite{HS01} and to the 
different deformation between proton and neutron, as we discuss below. 

In Ref. \cite{TFHB97}, Terasaki {\it et al.} have suggested that 
a low-lying quadrupole isovector mode 
may appear when the protons and the neutrons 
have different deformation parameters. In neutron-rich nuclei, the 
isovector and the isoscalar modes are coupled to each other \cite{HSZ97}, 
and thus this effect would also enhance the low-lying 
isoscalar quadrupole strength. In order to confirm this, we repeat the 
calculations by setting the proton and the neutron deformation parameters 
to be identical. Figure 2 shows the low-lying part of 
the isoscalar quadrupole response of $^{38}$Mg for the 
$K^\pi=0^+$ (the top panel), $K^\pi=1^+$ (the middle panel), 
and $K^\pi=2^+$ (the bottom panel). The dotted and the dashed 
curves are for the case where the proton and the neutron deformations are 
set to be the smaller value 0.28 and the larger value 0.33, respectively. 
The solid line denotes the original calculation shown in fig. 1, 
where the proton and the neutron have different deformations. 
For $K^\pi=0^+$ and 1$^+$, those three lines are very similar to each other, 
while, in marked contrast, 
the lowest $K^\pi=2^+$ excitation (the gamma vibration) 
clearly shows a different behaviour. 
When the deformation is different between proton 
and neutron, the excitation energy of this stae becomes significantly smaller 
and at the same time 
the strength is increased considerably. The effect of a larger 
proton deformation 
parameter simply changes the dotted curve to the dashed curve. 
There is an additional effect 
originating from the different proton and neutron deformations, 
that enhances the collectivity of the gamma vibration. 

\section{Concluding remarks}

At present, a continuum QRPA calculation based on the HFB method is 
still too costly to apply to 
deformed nuclei. The only possible way to overcome 
this difficulty is to employ  
the BCS approximation for the pairing correlations, where the particle-hole 
and the particle-particle channels are decoupled. Contrary to a general 
belief, the BCS approximation for the description of ground state 
properties has been shown adequate even for neutron-rich nuclei 
\cite{G01,SGL00}. 
We presume that this is the case also for the description of excited states, 
and have developed the QRPA method based on the BCS approximation. 
Particularly, we have presented the response function method for deformed 
QRPA, which substantially reduces 
the computational effort to solve QRPA compared with 
the conventional method which diagonalises the non-hermitian QRPA matrix. 
We have applied the method 
to quadrupole responses of the deformed neutron-rich 
nucleus $^{38}$Mg. We have shown that the collectivity of the gamma vibration 
is increased when the protons and the neutrons deform in a different way from 
each other as the mean-field calculation predicts. 
 
In this paper, we have discretized the continuum states by putting the nucleus 
in a large spherical box. For a more consistent description, one would need 
to treat the continuum spectra exactly. 
This applies both to the BCS and to the QRPA calculations. In the continuum 
BCS calculation for the ground state, the single-particle resonance states 
play an essential role \cite{SGL00}. 
Recently, we have shown that the coupled-channels approach with the 
eigenchannel basis offers an efficient way to identify a deformed 
single-particle resonance state through the so called eigenphase 
sum \cite{HG03}. 
We have also derived a factorization formula for the energy and the radial 
dependences of a scattering wave function near an isolated resonance 
state \cite{HG03}. This formula will be useful when one computes the matrix 
elements of pairing interaction over a wide energy region 
including the continuum 
spectra. 
As for a continuum QRPA calculation, 
it is well known that the exact treatment of the continuum couplings 
is possible if one uses the 
single-particle Green function in the coordinate space 
representation \cite{SB75}. 
An extension of this approach to a deformed system has been done in the field 
of atomic and molecular physics \cite{LS84}, which again uses 
the coupled-channels method with the eigenchannel basis 
(see also Ref. \cite{NY01}). 
We have also recently extended the Green function approach to spherical 
open-shell nuclei within the BCS approximation \cite{HS01}. 
We plan to combine these two extensions of the Green function approach 
in order to calculate the continuum response of superfluid deformed 
nuclei built 
on top of the continuum BCS ground state. We will report the results 
elsewhere. 


\begin{figure}[htb]
\begin{minipage}[t]{75mm}
\includegraphics[scale=0.45]{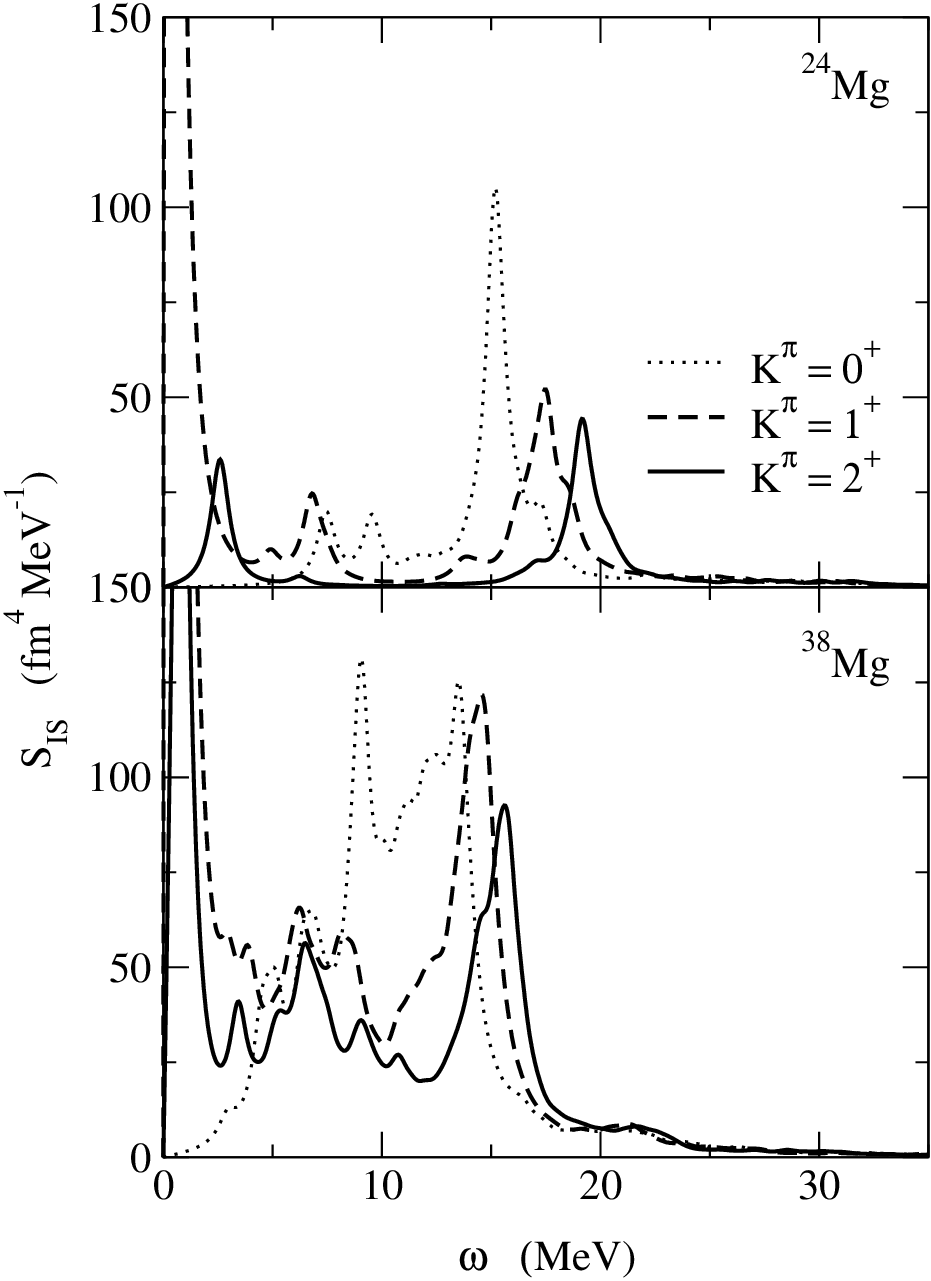}
\caption{The QRPA responses of the $^{24}$Mg nucleus 
(the upper panel) and of the $^{38}$Mg nucleus (the lower panel) 
to an isoscalar quadrupole field. The dotted, dashed, and solid 
lines are for the $K^{\pi}=0^+$, $1^+$, and $2^+$ modes, respectively.}
\end{minipage}
\hspace{\fill}
\begin{minipage}[t]{75mm}
\includegraphics[scale=0.48]{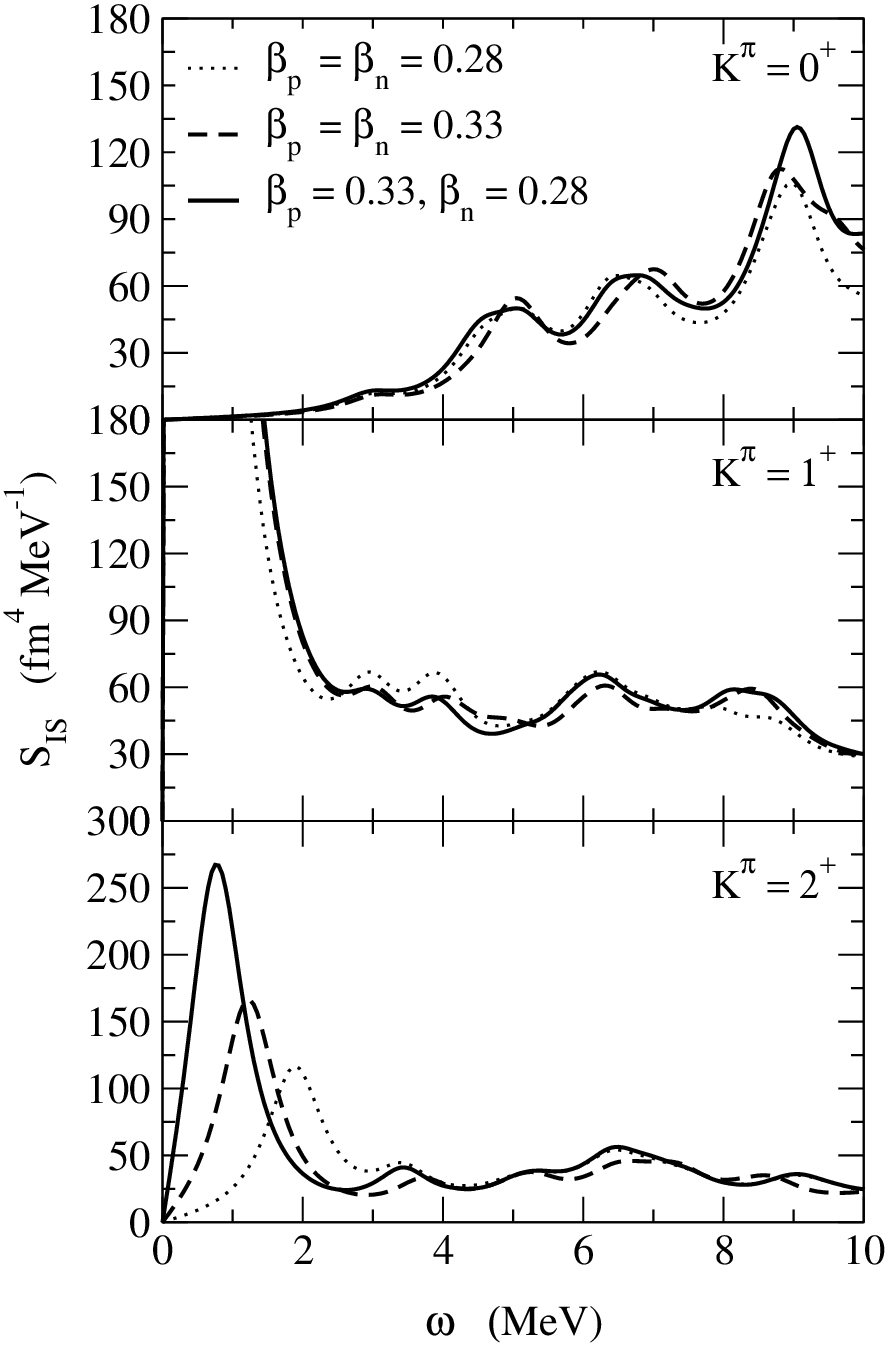}
\caption{The effect of different proton and neutron deformations on the 
isoscalar quadrupole response of the $^{38}$Mg nucleus. 
The dotted and the dashed lines are obtained by setting the proton and 
the neutron deformations to be the same as indicated in the inset, while 
the solid line is the result when the proton deformation is larger than the 
neutron deformation by 0.05. The top, middle, and bottom panels are 
for the $K^{\pi}=0^+$, $1^+$, and $2^+$ modes, respectively.}
\end{minipage}
\end{figure}


\begin{thebibliography}{9}

\bibitem{M02}T. Motobayashi, Eur. Phys. J. {\bf A15} (2002) 99. 

\bibitem{Y01}K. Yoneda {\it et al.}, Phys. Lett. {\bf B499} (2001) 233. 

\bibitem{I01}H. Iwasaki {\it et al.}, Phys. Lett. {\bf B522} (2001) 227. 

\bibitem{S02}
D. Sohler {\it et al.}, Phys. Rev. {\bf C66} (2002) 
054302. 

\bibitem{WBB90}E.K. Warburton, J.A. Becker, and B.A. Brown, 
Phys. Rev. C{\bf 41} (1990) 1147. 

\bibitem{O01}
T. Otsuka {\it et al.}, Phys. Rev. Lett. {\bf 87} (2001) 
082502. 

\bibitem{D96}J. Dobaczewski 
{\it et al.}, Phys. Rev. {\bf C53} (1996) 2809. 

\bibitem{G01}
M. Grasso 
{\it et al.}, Phys. Rev. {\bf C64} (2001) 064321. 

\bibitem{M01}M. Matsuo, Nucl. Phys. {\bf A696} (2001) 371. 

\bibitem{K02}
E. Khan 
{\it et al.}, Phys. Rev. {\bf C66} (2002) 024309. 

\bibitem{SGL00}N. Sandulescu, Nguyen Van Giai, and R.J. Liotta, 
Phys. Rev. {\bf C61} (2000) 061301(R). 

\bibitem{SM84}Y.R. Shimizu and K. Matsuyanagi, Prog. Theo. Phys. {\bf 72} 
(1984) 1017; T. Nakatsukasa, S. Mizutori, and K. Matsuyanagi, 
Prog. Theo. Phys. {\bf 87} (1992) 607. 

\bibitem{ZSP78}D. Zawischa, J. Speth, and D. Pal, Nucl. Phys. 
{\bf A311} (1978) 445. 

\bibitem{HS01}K. Hagino and H. Sagawa, Nucl. Phys. {\bf A695} (2001) 
82. 

\bibitem{HG03}K. Hagino and Nguyen Van Giai, e-print: nucl-th/0305034. 

\bibitem{J78}B.R. Johnson, J. Chem. Phys. {\bf 69} (1978) 4678. 

\bibitem{R68}D.J. Rowe, Nuclear Collective Motion, Methuen, London, 
1968. 

\bibitem{GSV98}Nguyen Van Giai, Ch. Stoyanov, and V.V. Voronov, 
Phys. Rev. {\bf C57} (1998) 1204. 

\bibitem{BM69}A. Bohr and B.R. Mottelson, 
Nuclear Structure Vol. 1, Benjamin, New York, 1969. 

\bibitem{TFHB97}J. Terasaki, H. Flocard, P.-H. Heenen, and P. Bonche, 
Nucl. Phys. {\bf A621} (1997) 706. 

\bibitem{S3}M. Beiner, H. Flocard, N. Van Giai, and P. Quentin, 
Nucl. Phys. {\bf A238} (1975) 29. 

\bibitem{YLC99}D.H. Youngblood, Y.-W. Lui, and H.L. Clark, 
Phys. Rev. {\bf C60} (1999) 067302, and references therein. 

%

\bibitem{HSZ97}I. Hamamoto, H. Sagawa, and X.Z. Zhang, 
Phys. Rev. {\bf C55} (1997) 2361. 

\bibitem{SB75}S. Shlomo and G.F. Bertsch, Nucl. Phys. {\bf A243} (1975) 507. 

\bibitem{LS84}Z.H. Levine and P. Soven, 
Phys. Rev. {\bf A29} (1984) 625. 

\bibitem{NY01}T. Nakatsukasa and K. Yabana, J. Chem. Phys. {\bf 114} 
(2001) 2550. 

\end{thebibliography}
\end{document}